\documentclass{kluwer}    
\usepackage{graphicx}

\newdisplay{guess}{Conjecture}

\begin{document}
\begin{article}
\begin{opening}
\title{{\bf The Role of Magnetic Fields in Spiral Galaxies}}
\author{Rainer \surname{Beck}}
\runningauthor{Rainer Beck}
\runningtitle{The Role of Magnetic Fields}
\institute{MPI f\"ur Radioastronomie, Auf dem H\"ugel 69,
53121 Bonn, Germany}

\begin{abstract}
Interstellar magnetic fields are strong: up to 25$\mu$G in spiral
arms and 40$\mu$G in nuclear regions.
In the spiral galaxy NGC~6946 the average magnetic energy density
exceeds that of the thermal gas. Magnetic fields control the evolution of dense
clouds and possibly the global star formation efficiency in galaxies.
Gas flows and shocks in spiral arms and bars are modified by magnetic
fields. Magnetic forces in
star-forming circumnuclear regions are able to drive mass inflow towards
the active nucleus. Magnetic fields are essential for the propagation
of cosmic rays and the formation of galactic winds and halos.
\end{abstract}

\end{opening}

\section{Interstellar Magnetic Fields Are Hard To Observe}

{\it ... but it's easier than most people think.}
In radio continuum the typical degrees of polarization are much higher
than in the other spectral ranges, and we benefit from the development
of large instruments and sensitive receivers. This is why most of our
knowledge on interstellar magnetic fields in our Galaxy
(\opencite{Reich1994}; \opencite{Heiles1996}; \opencite{Beck2001})
and in external galaxies is based on polarized radio emission and 
its Faraday rotation (see also \opencite{Beck2000}; \opencite{Beck2002}). 
To detect extended features and achieve high resolution,
data from interferometric (synthesis) and single dish telescopes 
have to be combined.

\section{Radio Waves Are Tracers of Magnetic Fields}

{\it ... simply the best ones.}
Interstellar magnetic fields are illuminated by cosmic-ray electrons
emitting synchrotron radiation,
the dominant contribution to the diffuse radio continuum emission at
centimeter and decimeter wavelengths. Synchrotron emission is
intrinsically highly linearly polarized, 70--75\% in a completely
regular magnetic field. The observable degree of polarization in
galaxies is reduced by a contribution of unpolarized thermal emission
which may dominate in star-forming regions, by Faraday depolarization 
(\opencite{Sokoloffetal1998}) and by geometrical depolarization within the beam.
A map of the {\it total} nonthermal intensity (Fig.~1) reveals the
{\it strength of the total interstellar magnetic fields} in the plane of
the sky, while the {\it polarized intensity and polarization angle} (Figs.~2 and 3)
reveal the {\it strength and structure of the resolved regular fields}
in the plane of the sky. 
The orientation of polarization vectors is changed in a magneto-ionic 
medium by {\it Faraday rotation} which is generally small
below about $\lambda6$~cm so that the $\mathbf{B}$--vectors (i.e. the
observed $\mathbf{E}$--vectors rotated by $90^{\circ}$) directly trace
the {\it orientation\/} of the regular fields in the sky plane.

\begin{figure}[htb]
\center
\includegraphics[bb = 51 95 555 735,width=9cm]{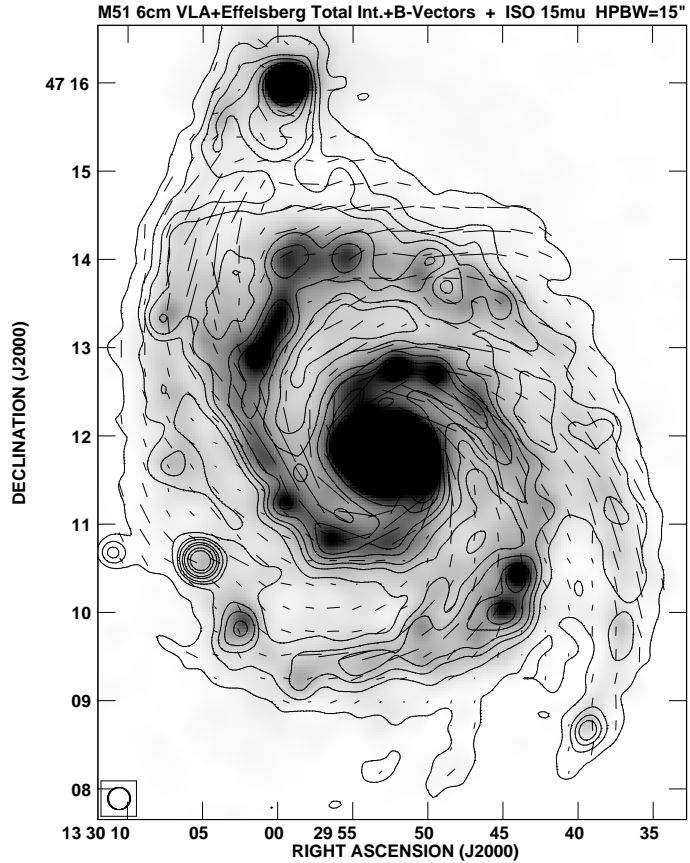}
\caption{ Total radio continuum intensity (contours) and 
$\mathbf{B}$--vectors of polarized intensity of M\,51 at 
$15^{\prime\prime}$ resolution, combined from 
VLA and Effelsberg observations at $\lambda$6~cm. The grey-scale 
ISOCAM image shows the $\lambda 15\, \mu$m emission 
(kindly provided by M.~Sauvage), smoothed to $15^{\prime\prime}$ 
resolution. (Beck, unpublished)}
\label{fig:m51}
\end{figure}

\section{Magnetic fields {\it Are} Strong}

{\it ... much stronger than most people think.}
The average strength of the total $\langle B_{\mathrm{t},\perp}\rangle$
and the resolved regular field $\langle B_{\mathrm{reg},\perp}\rangle$
in the plane of the sky can be derived from the total and polarized
radio synchrotron intensity, respectively, if energy-density {\it
equipartition} between cosmic rays and magnetic fields is assumed
(\opencite{Becketal1996}). 

In our Galaxy the accuracy of the equipartition assumption can be
tested, because we have independent information from in-situ measurements 
about the local cosmic-ray energy density and from $\gamma$-ray data
about their distribution.
Combination of the radio synchrotron emission, the local cosmic-ray
electron density and diffuse continuum $\gamma$-rays yields a local
strength of the total field of $6\pm1\,\mu$G
(\opencite{Strongetal2000}), the same value as derived from
energy equipartition (Berkhuijsen in \opencite{Beck2001}). 
Even the scale length of the radial variation of the equipartition 
field strength is similar to that in \inlinecite{Strongetal2000}.

The mean equipartition strength of the total field for a sample of 
74 spiral galaxies (derived from their total radio flux densities) is 
$\langle B_\mathrm{t}\rangle = 9\,\mu$G, 
ranging between $\langle B_\mathrm{t}\rangle \simeq 6\,\mu$G
in radio-faint galaxies like M\,31 and M\,33, and $\simeq
15\,\mu$G in grand-design galaxies like M\,51, M\,83 and NGC~6946. In
prominent spiral arms the total field strength is 20--25$\mu$G,
up to 40$\mu$G in nuclear regions (Sect.~7.3), and even
1~mG in molecular clouds compressed by
supernova shocks (\opencite{Broganetal2000}), in compact cloud cores
(\opencite{MyersandGoodman1988}; \opencite{Uchidaetal2001})
and in filaments near the Galactic center (\opencite{Reich1994};
\opencite{Yusef-Zadehetal1996}).

\section{Magnetic Fields {\it Are} Dynamically Important}

{\it ... more important than most people think.}
A standard approach to estimate the dynamical importance of various
competing forces is to compare the corresponding energy densities (Table~1).
Note, however, that the dynamical effects of magnetic fields are 
anisotropic due to their vector nature.
NGC~6946 (Fig.~2) is a useful standard spiral galaxy because it has no
companion, no strong density waves, no active nucleus and has been observed 
in many spectral ranges. The average total equipartition field strength 
in the inner disk (within 4~kpc radius) is $B_t\simeq 18\,\mu$G (using the
map of synchrotron emission by \opencite{Walshetal2002}) and the
average density of the warm ionized gas ($T\simeq 10^4$~K) is $n_e\simeq
0.5~\mathrm{cm}^{-3}$ (using their map of thermal radio emission and a
filling factor of 5\%). The map of the total neutral gas (molecular +
atomic, where the molecular gas dominates until about 5~kpc radius) 
yields a mean number density (within a disk of 4~kpc radius 
and 100~pc height) of $n\simeq 20$~H~atoms $\mathrm{cm}^{-3}$. 
The turbulent velocity of the neutral 
gas is assumed to be $v_\mathrm{turb} = 7.5$~km/s, as for the cold
neutral gas in our Galaxy (\opencite{KalberlaandKerp1998}), and
the global rotation velocity $v_\mathrm{rot} = 170$~km/s. For the
cosmic rays we know the slope of the energy spectrum
($N[E]\propto N_0 E^{-2\alpha -1} dE$) from the observed synchrotron
spectral index $\alpha$, but no information on the absolute particle
number $N_0$ is available so that we have to rely on the validity of
energy equipartition with the magnetic fields.

\begin{table}
\caption{Average energy densities $\epsilon$ (in $\rm 10^{-12}\, erg\, cm^{-3}$)
in the ISM of the inner disk ($\le$4~kpc radius) of NGC~6946}
\begin{tabular}{lllr}
\hline\noalign{\smallskip}
\noalign{\smallskip}
Magnetic field (equip.)   &$\epsilon_{B_\mathrm{t}^2}$& $B^2 / 8 \pi$ 
                 &13 \\
Cosmic rays (equip.)      &$\epsilon_\mathrm{CR}$& $C \int E^{-2\alpha} dE$ 
                 &13 \\
Warm ionized medium &$\epsilon_\mathrm{WIM}$& $\frac{3}{2}\, n_e\, k\, T$ 
                 &1 \\
Turbulent cloud motion &$\epsilon_\mathrm{turb}$& $\frac{1}{2}\, \rho\, v_\mathrm{turb}^2$  
                 &10 \\
Global gas rotation &$\epsilon_\mathrm{rot}$& $\frac{1}{2}\, \rho\, v_\mathrm{rot}^2$ 
                 &5000 \\
\noalign{\smallskip}
\hline
\end{tabular}
\end{table}

The ratio $\beta$ between the thermal and magnetic energy densities 
of the warm ionized medium (WIM) is only $\simeq0.1$. 
The hot ionized gas and the warm neutral gas are believed to be
in pressure equilibrium with the WIM so that their $\beta$ should be 
similar. (The thermal energy of the cold gas is negligible.) 
A low thermal contribution to the
total gas pressure was also found in the local Galactic ISM 
(\opencite{JenkinsandTripp2001}; Jenkins, this volume). A
significant fraction of the diffuse ISM must be unstable, giving rise
to gas flows. 
From X-ray and radio continuum observations in the halo of the spiral
galaxy M~83, \inlinecite{Ehleetal1998} derived a similarly low value of
$\beta$. The magnetic field dominates thermal processes in the disk 
and in the halo of galaxies.

\section{Magnetic Fields Drive Star Formation}

Turbulent motion of neutral gas and total magnetic fields in the inner
disk of NGC~6946 are roughly in energy equipartition (Table~1)
(but see Sect.~6). This fact supports
the idea that magnetic fields are anchored in the (partly ionized)
envelopes of gas clouds (\opencite{Beck1991}). This also explains
the tight radio--infrared correlation (\opencite{NiklasandBeck1997};
\opencite{Walshetal2002}). Between the clouds, field loops can be generated 
by the Parker instability which can drive
the galactic dynamo (\opencite{HanaszandLesch1998};
\opencite{Mossetal1999}; Hanasz et al., this volume).

Interstellar magnetic fields affect the motion of small gas 
clouds (\opencite{Elmegreen1981}) and their collision rate. The 
importance of internal fields for the evolution of
clouds is generally accepted,
though not understood (\opencite{MestelandParis1984};
\opencite{MyersandGoodman1988}; \opencite{Crutcher1999};
\opencite{Heitschetal2001}). However, little is known
about the influence of magnetic fields in the ISM on general properties
of star formation.

Stability of cloud cores depends also on the fractional ionization by
cosmic rays, and the number density of cosmic rays in the ISM is
controlled by the interstellar magnetic fields. A larger fractional
ionization allows a better coupling of the fields to the gas (weaker
ambipolar diffusion) and thus stronger internal fields. As a result,
the field strength inside and outside of the clouds should be
correlated.

The star formation rate $SFR$ is known to relate non-linearly with
gas density $\rho$, the {\it Schmidt law} $SFR\propto \rho^\mathrm{n}$,
with $n\simeq 1.5$ (\opencite{WongandBlitz2002}). Hence, the
star formation efficiency ($SFE \propto SFR/\rho$) increases with gas
density. As the total field strength $B_t$ in galaxies is known to
scale globally with $\rho$ as $B_t\propto \rho^{0.5}$
(\opencite{NiklasandBeck1997}), a relation $SFE\propto B_t^\mathrm{x}$
could well be the cause of the non-linearity. The observational data
are still poor (\opencite{Vallee1994}). The supernova-driven ISM model
including magnetic fields by \inlinecite{GazolandPassot1999} indeed
predicts a significant increase of the average star-formation rate with
field strength. Furthermore, strong fields may shift the stellar mass
spectrum towards the more massive stars (\opencite{Mestel1994}).

\section{Do Magnetic Fields Dominate in Outer Galaxies ?}

Most observable quantities in galactic disks have exponential 
distributions. For NGC~6946 the scale length of the synchrotron 
emission is $l_\mathrm{syn}=3.9\pm0.1$~kpc (radial range 3--8~kpc). 
The synchrotron disk is a combination of the distributions of cosmic-ray
electrons and total magnetic fields.
In case of equipartition between these two
components, the scale length of the total magnetic field is
$l_\mathrm{B} = (3+\alpha) l_\mathrm{syn} \simeq 4 l_\mathrm{syn}
\simeq$16~kpc, and the scale lengths of the magnetic energy density 
and of the cosmic rays are $l_\mathrm{B^2}=l_\mathrm{CR}\simeq\,${\bf 8~kpc}. 
All other energy densities in Table~1 fall off with the 
scale length of the neutral gas $l_{\rho}=3.2\pm$0.1~kpc if
$T$, $v_\mathrm{turb}$ and $v_\mathrm{rot}$ are radially constant,
or with a smaller scale length if one of these parameters
decreases with radius. 

The scale length of the distribution 
of supernova remnants (the sources of cosmic rays) in NGC~6946 is 
only $\simeq$2~kpc (Sasaki et al., this volume) so that a large
diffusion length is needed to reach equipartition also at large
radii (see the cosmic-ray propagation model by 
\opencite{Breitschwerdtetal2002}). A large value of 
$l_\mathrm{CR}$ is also required to explain the distribution of 
Galactic $\gamma$-rays (\opencite{Strongetal2000}).

As a consequence of the different scale lengths
$\beta=\epsilon_\mathrm{turb}/\epsilon_\mathrm{B^2}$ decreases with
radius: The dominance of the magnetic field increases towards large radii.
It is however hard to understand why the magnetic field energy density is 
larger that of the turbulent motions. This can be in conflict with
turbulent generation of interstellar magnetic fields which predicts
energy equipartition between turbulent motions and magnetic fields, and 
hence $l_\mathrm{B^2} = l_{\rho}$ (if $v_\mathrm{turb}\simeq\,$const) and 
$l_\mathrm{syn} < l_{\rho}$. Radial diffusion 
of the magnetic field (\opencite{Priklonskyetal2000}), field connections
through the wind-driven halo (\opencite{Breitschwerdtetal2002}) 
or a supra-equipartition turbulent dynamo (\opencite{Belyaninetal1993})
are possible explanations. 

In the outermost parts of galaxies the magnetic field energy density 
may even reach the level of global rotational gas motion. If so, the 
rotation curves at large radii could be affected by magnetic fields, 
as discussed by \inlinecite{BattanerandFlorido2000}.
Field strengths in the outer parts of galaxies are difficult to measure.
Synchrotron emission is weak, but Faraday rotation of polarized 
background sources can be measured out to large radii. 
\inlinecite{Hanetal1998} found evidence for regular fields in M\,31
at 25~kpc radius of similar strength as in the inner disk.
More detailed studies in a number of galaxies
are required.

\section{Magnetic Fields Act Everywhere}

\subsection{Spiral Arms}

Maps of the total radio emission and ISOCAM maps
of the mid-infrared dust emission (Fig.~1) reveal a surprisingly close 
connection (\opencite{Fricketal2001}; \opencite{Walshetal2002}),
suggesting a coupling of the {\it total} magnetic field to the warm 
dust mixed with cool gas ($B_t\propto \rho^{0.5}$, 
\opencite{NiklasandBeck1997}). Strongest total
fields generally coincide with highest emission from dust and 
gas in the spiral arms (Fig.~1), whereas the polarized emission is
weak in the arms due to field tangling by turbulent motions
or by supernova shock fronts. Radio polarization observations show that
in most galaxies the {\it regular} field follows the spiral structure
seen in the stars and the gas, though generally offset, sometimes
forming {\it magnetic spiral arms} between the gaseous arms
(Fig.~2). Here the energy density of the regular field alone 
($\simeq 10\,\mu$G) is comparable to that of the neutral interarm gas 
($\rho \simeq 5 \mathrm{cm}^{-3}$).

In galaxies with strong density waves (like M~51) the regular field is
strong at the positions of the dust lanes on the inner edge
of the spiral arms. However, the arm-interarm contrast of the regular
field is low. The strong field may modify the gas flow and the
compression ratio.

The observation of large-scale patterns in Faraday rotation measures
proves that the regular field in galaxies has a
coherent direction and is not generated by compression or stretching of irregular
fields in gas flows. The {\it dynamo} mechanism is able to generate and
preserve coherent magnetic fields of spiral shape
(\opencite{Becketal1996}; \opencite{Mossetal1999}). The observed
alignment of magnetic pitch angles with those of the gaseous arms
can be achieved by inclusion of the shear flow around spiral arms 
(\opencite{Lindenetal1998}). Future dynamo models should address the
interaction between dynamo wave and density wave.

\begin{figure}[htb]
\center
\includegraphics[bb = 40 191 568 639,width=9cm]{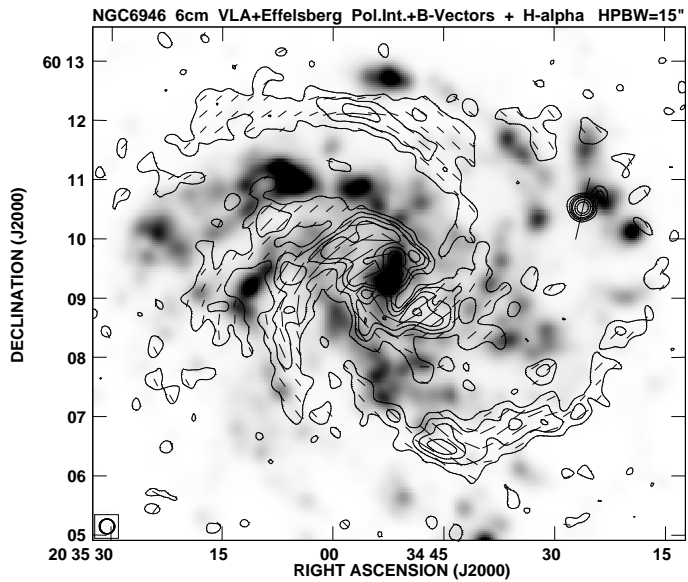}
\caption{Polarized radio intensity (contours) and 
$\mathbf{B}$--vectors of polarized 
intensity of NGC~6946 at $15^{\prime\prime}$ resolution, combined 
from VLA and Effelsberg observations at $\lambda$6~cm. The grey-scale 
image shows the H$\alpha$ emission (kindly provided by A.~Ferguson), 
smoothed to $15^{\prime\prime}$ resolution. 
(From \protect\opencite{BeckandHoernes1996})
}
\label{fig:ngc6946}
\end{figure}

\subsection{Bars}

Average radio intensity, radio luminosity and star-formation activity
in barred galaxies correlate with {\it relative bar length}
(\opencite{Becketal2002}). In the best-studied case, NGC~1097, the
general similarity of the $B$--vectors (Fig.~3) and gas streamlines around
the bar as obtained in simulations is striking (\opencite{Becketal1999}).
The field lines are a tracer of the sheared flow pattern in the sky
plane which is otherwise unobservable. Remarkably, the optical
image of NGC~1097 shows dust filaments outside the bar which are
aligned with the field.

The bar's major dust lane is believed to indicate the shock front.
However, the region of strongest shear in NGC~1097 (where the field
changes its direction abruptly) is located $\simeq 1$~kpc in front
of the dust lane (Fig.~3), in conflict with the hydrodynamical models.
Furthermore, the regular field is only weakly compressed in the
bar. Either the field decouples from the gas (at least from its
diffuse component) and thus avoids the gas shock, or there is no shock
at all. Detailed observations of the velocity field are required.
\inlinecite{Mossetal2001} used velocity shear and dynamo action 
to model the field configuration, but a MHD model including the
back-reaction of the field onto the gas flow is still needed.

\begin{figure}[htb]
\center
\includegraphics[bb = 40 112 570 717,width=9cm]{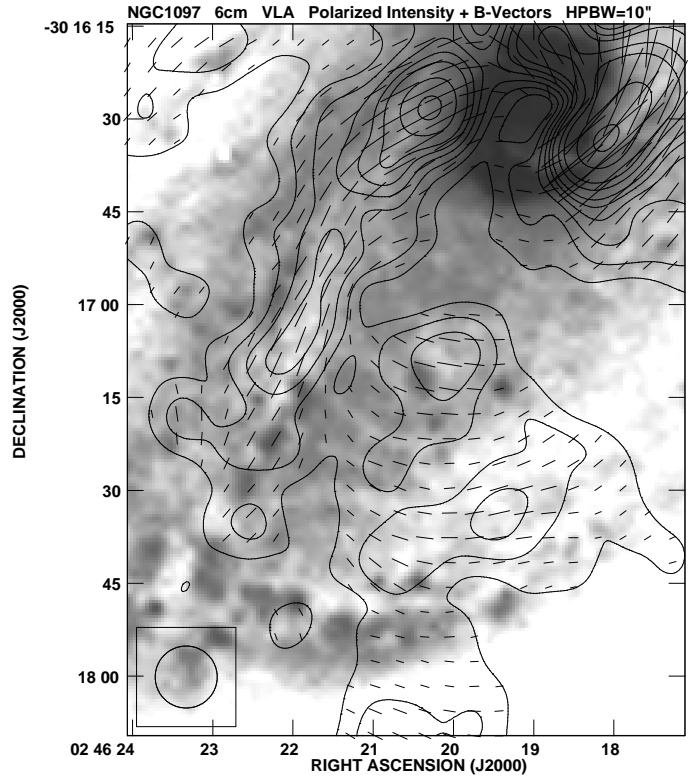}
\caption{Polarized radio intensity of NGC~1097 (contours) and 
$\mathbf{B}$--vectors 
of polarized intensity at $10^{\prime\prime}$ resolution, observed 
with the VLA at $\lambda$6~cm. The grey-scale image shows an
optical image from the Cerro Tololo Observatory (kindly provided 
by H.~Arp). (From \protect\opencite{Mossetal2001})}
\label{fig:ngc1097}
\end{figure}

\subsection{Nuclear Regions}

Many barred galaxies have circumnuclear rings, sites of ongoing
intense star formation, sometimes with an active nucleus in the center.
Radio polarization maps of these inner regions (\opencite{Becketal2002})
revealed strong regular fields (up to 40$\mu$G) 
with spiral patterns and large pitch
angles, accompanied by spiral dust filaments visible on optical images.
Magnetic stress in the circumnuclear ring can drive mass inflow 
to feed the active nucleus, as suggested for NGC~1097
(\opencite{Becketal1999}).

\subsection{Halos}

In edge-on galaxies the observed field orientations are mainly parallel
to the disk (\opencite{Dumkeetal1995}). A prominent exception is
NGC~4631 with the brightest and largest halo observed so far
(\opencite{Beck2000}), composed of vertical magnetic spurs connected to
star-forming regions in the disk (\opencite{GollaandHummel1994}). The
field is probably dragged out by a strong galactic wind. The magnetic
energy density in the halo of, e.g. M\,83, exceeds that of the hot gas
(\opencite{Ehleetal1998}). Halo magnetic fields are important for the
propagation of cosmic rays, the formation of a galactic wind
(\opencite{Breitschwerdtetal2002}) and the stability of gas filaments
(\opencite{Tuellmannetal2000}; Dettmar, this volume).

\section{Outlook}

Much has been learnt about the global properties of interstellar
magnetic fields, but their detailed structure on scales below about 100~pc
is still unclear. Polarization observations in our Galaxy trace
structures on pc and sub-pc scales (e.g. \opencite{Uyanikeretal1999};
\opencite{Gaensleretal2001}). 
Future radio polarization observations with increased sensitivity
and resolution (Extended VLA, Square Kilometer Array) will show the
full wealth of magnetic structures in galaxies.

\end{article}
\end{document}